 \def\Title{Compatibility and probability}
 \def\arXiv{quant-ph/0403021}
 \def\Abstract{\normalsize%
A review of various definitions of ``compatibility'' expressed in terms of ordinary probability, and a discussion of the occurrence of incompatibility (and the related phenomenon of interference) in non-quantal probabilistic systems.
 }%
\documentclass[aps,rmp,nobibnotes,nofootinbib,onecolumn,twoside,noshowpacs,draft]{revtex4}
 \usepackage[tbtags,sumlimits]{amsmath}
 \usepackage{amssymb}
 \makeatletter
 \def\p@section{}
 \def\p@subsection{}
 \def\p@subsubsection{}
 \def\p@paragraph{}
 \def\p@subparagraph{}

 \def\@hangfrom@section#1#2#3{\@hangfrom{#1}{\large\textrm{#2}}{\large\textrm{#3}}}%
 \def\@hangfrom@subsection#1#2#3{\@hangfrom{#1}{\textrm{#2}}{\textrm{#3}}}%
 \def\@hangfrom@subsubsection#1#2#3{\@hangfrom{#1}{\textrm{#2}}{\textrm{#3}}}%
 \def\frontmatter@setup{\normalfont\raggedright}% Not san serif now
 \makeatother

 \usepackage{xspace}

 \newcommand{\ie}{i.e., }
 \newcommand{\eg}{e.g., }

 \newcommand{\vN}{von Neumann\xspace}

 \newcommand{\Luders}{L{\"u}ders\xspace}
 \newcommand{\RefSec}[1]{Sec.~\ref{#1}}
 
 \newcommand{\IFF}{\/i\/f\/f\/\xspace}
 \newcommand{\set}[1]{\ensuremath{{\left\{\,#1\,\right\}}}}%

 \newcommand{\Orj}[2]{\ensuremath{{\textstyle\bigvee}_{\!#1}\,{#2}}}
 \newcommand{\andthen}{\ensuremath{\,\&\,}}
 \newcommand{\KDelta}[2]{\ensuremath{\delta_{{#1}{#2}}}}
 \renewcommand{\Pr}[2][]{\ensuremath{{\rm Pr}_{#1}\bigl(\,{#2}\,\bigr)}}
 \newcommand{\Prob}[3][]{\ensuremath{{\rm Pr}_{#1}\bigl(\,{#2}\bigm|#3\,\bigr)}}
 \newcommand{\Sys}[1]{\ensuremath{\mathcal{#1}}\xspace}
 
 \newcommand{\ProjSym}{\boldsymbol{\mathsf{P}}}
 \newcommand{\Proj}[2][]{\ensuremath{\ProjSym^{\Sys{#1}}[\,#2\,]}\xspace}
 \newcommand{\bRho}{\pmb{\rho}}
 
 \newcommand{\Trace}[2][]{\ensuremath{{\rm Tr}%
                    {\!}_{\Sys{#1}}\left\{\,{#2}\right\}}}
 \usepackage{amsthm}
 \newtheorem{theorem}{Theorem}

 \newtheorem*{theorem*}{Theorem}
 \newtheorem*{lemma*}{Lemma}
 \newtheorem*{postulate*}{Postulate}
 \theoremstyle{definition}
 
 \newtheorem*{definition*}{Definition}

 \newcommand{\p}[1]{\ensuremath{p_{#1}}\xspace}
 \newcommand{\q}[1]{\ensuremath{q_{#1}}\xspace}
\newcommand{\ProjPj}{\boldsymbol{P}_j}
\newcommand{\ProjPjp}{\boldsymbol{P}_{j'}}
\newcommand{\ProjQ}{\boldsymbol{Q}}
\newcommand{\ProjQk}{\boldsymbol{Q}_k}
\newcommand{\ProjC}{\boldsymbol{C}}

 \newcommand{\Color}{\textsl{Color}\xspace}
 \newcommand{\Pattern}{\textsl{Pattern}\xspace}

\begin{document}
 \makeatletter
 \def\ps@titlepage{%
   \renewcommand{\@oddfoot}{}%
   \renewcommand{\@evenfoot}{}%
   \renewcommand{\@oddhead}{\hfill\arXiv}
   \renewcommand{\@evenhead}{}}
 \makeatother
%%%%%%%%%%%%%%%%%%%%%%%%%%%%%%%%%%%%%%%%%%%%%%%%%%%%%%%%%%%%%%%%%%%%%%%%%%%%%%%%%%%%%%%
\title[Kirkpatrick -- \Title] %% for running titles on pages
      {\Title} %% the title-page title
%%%%%% PERSONAL %%%%%%%%%%%%%%%%%%%%%%%%%%%%%%%%%%%%%%%%%%%%%%%%%%%%%%%%%%%%%%%%%%%%%%%%
\author{K.~A.~Kirkpatrick}
\email[E-mail: ]{kirkpatrick@physics.nmhu.edu}
\affiliation{New Mexico Highlands University, Las Vegas, New Mexico 87701}
%%%%%%%%%%%%%%%%%%%%%%%%%%%%%%%%%%%%%%%%%%%%%%%%%%%%%%%%%%%%%%%%%%%%%%%%%%%%%%%%%%%%%%%%
%%%%%% ABSTRACT %%%%%%%%%%%%%%%%%%%%%%%%%%%%%%%%%%%%%%%%%%%%%%%%%%%%%%%%%%%%%%%%%%%%%%%%
\begin{abstract}
 \Abstract
\end{abstract}
%%%%%%%%%%%%%%%%%%%%%%%%%%%%%%%%%%%%%%%%%%%%%%%%%%%%%%%%%%%%%%%%%%%%%%%%%%%%%%%%%%%%%%%%
 \maketitle
%%%%%%%%%%%%%%%%%%%%%%%%%%%%%%%%%%%%%%%%%%%%%%%%%%%%%%%%%%%%%%%%%%%%%%%%%%%%%%%%%%%%%%%%
%% centered short title in each header:
 \makeatletter\markboth{\hfill\@shorttitle\hfill}{\hfill\@shorttitle\hfill}\makeatother
 \pagestyle{myheadings}
%%%%%%%%%%%%%%%%%%%%%%%%%%%%%%%%%%%%%%%%%%%%%%%%%%%%%%%%%%%%%%%%%%%%%%%%%%%%%%%%%%%%%%%%
%%% BODY OF DOCUMENT %%%%%%%%%%%%%%%%%%%%%%%%%%%%%%%%%%%%%%%%%%%%%%%%%%%%%%%%%%%%%%%%%%%
%%%%%%%%%%%%%%%%%%%%%%%%%%%%%%%%%%%%%%%%%%%%%%%%%%%%%%%%%%%%%%%%%%%%%%%%%%%%%%%%%%%%%%%

\section{Introduction}

The notion of ``incompatibility'' of system variables arose with quantum mechanics; it is inconceivable in classical physics. Two expressions of incompatibility from the earliest days of quantum theory are the uncertainty principle of Heisenberg (measurement of one variable causes uncontrollable disturbance in other variables) and Bohr's ``complementarity'' (not all variables have simultaneous ``reality''). Interference --- the non-additivity of probabilities of disjoint alternatives --- arises only when a preparation of indistinguishable alternative values is followed by observation of a variable incompatible with those values.

In quantum mechanics, incompatibility is expressed by the non-commutativity of the operators corresponding to the variables; very little emphasis has been placed on the expressibility --- or meaning --- of incompatibility in classical terms.  Of course, variables in classical (deterministic dynamics) physics must be compatible --- but variables of systems which obey a classically probabilistic dynamics may very well be incompatible, in the sense that they satisfy a probability expression for incompatibility --- a probabilistic formula which, in the quantum setting, is equivalent to the non-commutativity of the corresponding operators. Not only have such expressions been known since \Luders' \citeyear{Luders51} paper (at the latest), we now \citep{Kirkpatrick:Quantal,Kirkpatrick:ThreeBox} have explicit examples of classical systems with incompatible variables --- incompatibility is not an exclusively quantal phenomenon.

After a very brief review of necessary (and well-known) matters of probability and quantum mechanics (\RefSec{S:Background}), we present in \RefSec{S:Expressions} improved treatments of two classical expressions of compatibility due to \Luders and one more due to \citet{Davies76}. In \RefSec{S:Definition}, considering all these, we present a favored classical definition of compatibility, and, in \RefSec{S:Example}, we discuss a classical example of incompatibility which exhibits interference.

\section{Notation and background}\label{S:Background}

The system has (at least) two variables $P$ and $Q$ with values \set{\p{j}} and \set{\q{k}}, respectively. The system passes through a sequence of ``measurements,'' at each of which one of the variables takes on a value randomly, $P=\p{j}$, say. Because we must consider sequences of several events, we introduce the simplifying notation for the conjunction of two successive event propositions, \andthen (``and then''), so the event $P=\p{j}$ followed by the event $Q=\q{k}$ is denoted $\p{j}\andthen\q{k}$ (``\p{j} and then \q{k}''). (In general, we abbreviate the proposition $P=\p{j}$ with the value \p{j}.)

The preparation of the system defines the probabilities of events; we call this state of affairs (and the implied equivalence class of preparations) the preparation-, or probability-, or more simply, p-state. Having prepared the system in the p-state $\sigma$, the probability in a measurement of $P$ that $P=\p{j}$ is written \Pr[\sigma]{\p{j}}.

Given the occurrence of $Q=\q{k}$ in a prior measurement of $Q$ following preparation of the p-state $\sigma$, the probability of \p{j} is denoted \Prob[\sigma]{\p{j}}{\q{k}}; \q{k}, the \emph{condition}, is equivalent to a \q{k}-filter in the preparation: $\Prob[\sigma]{\p{j}}{\q{k}}=\Pr[\sigma\andthen\q{k}]{\p{j}}$.  This conditional probability satisfies $\Prob[\sigma]{\p{j}}{\q{k}}%
  =\dfrac{\Pr[\sigma]{\q{k}\andthen\p{j}}}{\Pr[\sigma]{\q{k}}}$.

In quantum mechanics, to each elementary proposition $X=x$ corresponds a projector \Proj{x} (\Proj{x} is a 1-projector only if $x$ is an \emph{atomic} (non-degenerate) value); to each p-state $\sigma$ corresponds a trace-1 operator $\bRho[\sigma]$. The probability \Pr[\sigma]{\p{j}} is expressed in quantum terms as $\Trace{\bRho[\sigma]\,\Proj{\p{j}}}$. The conditional probability \Prob[\sigma]{y}{x} may be written 
$\Trace{\bRho[\sigma\andthen x]\Proj{y}}$, where 
$\bRho[\sigma\andthen x]=\dfrac{\Proj{x}\,\bRho[\sigma]\,\Proj{x}}
                               {\Trace{\bRho[\sigma]\,\Proj{x}}}$ \citep{Luders51}.
(For the case that $x$ is not atomic, \Luders' expression implicitly assumes Wigner's \emph{morality} --- essentially that there be no relative phase-shifts among the degenerate kets.)

\section{Classical expressions which lead to commutativity in quantum mechanics}%
\label{S:Expressions}

We present three criteria for compatibility expressed in terms of classical probability of occurrence of values of the variables. The first two, due to \citet{Luders51}, are clearly expressed in terms of \emph{disturbance}: is an established value of a variable disturbed by the occurrence of a value of another variable? is the probability of occurrence of a variable's value changed by a preceding measurement of another variable? The third, due to \citet{Davies76} and the author, is expressed in terms of the temporal reversal of measurement results: is the probability of $p$ followed by $q$ the same as the probability of $q$ followed by $p$? Each of these criteria is shown in the following theorems to lead, in the setting of quantum mechanics, to the commutability of the corresponding operators; the converse of each of these three theorems is true (with obvious, so not explicit, proof). Thus each criterion is \emph{equivalent} with commutability --- in the quantum setting, each is a definition of compatibility. 

To make the notation less awkward in the proofs of these three theorems, we will abbreviate \Proj{\p{j}} and \Proj{\q{k}} as  $\ProjPj$ and $\ProjQk$, respectively.

\subsection{\Luders' expressions of compatibility}

The first two criteria, and the corresponding theorems, are from \citet{Luders51} .

1. If a measurement of $Q$ resulting in the value \q{k} does not disturb the value \p{j} obtained in a previous measurement of $P$, then the corresponding projectors commute (the values \p{j} and \q{k} are ``compatible''):
\begin{theorem}[Non-disturbing measurement]\label{T:Luders1}
\begin{equation}
 \Prob[\sigma]{\p{j}}{\p{j}\andthen\q{k}}=1\;\forall\sigma%
 \;\Longrightarrow\;\Proj{\p{j}}\Proj{\q{k}}=\Proj{\q{k}}\Proj{\p{j}}.
\end{equation}
\end{theorem}
\begin{proof}: We have $\Prob[\sigma]{\p{j'}}{\p{j}\andthen\q{k}}= \Pr[\sigma\andthen\p{j}\andthen\q{k}]{\p{j'}}=\KDelta{j'}{j}$ (the last because the \set{\p{j}} are disjoint and complete), which in quantal terms is \[\Trace{\dfrac{\ProjQk\ProjPj\,\bRho[\sigma]\,\ProjPj\ProjQk}%
 {\Trace{\ProjPj\,\bRho[\sigma]\,\ProjPj\ProjQk}}\ProjPjp}=\KDelta{j'}{j};\] because this must hold for all $\bRho$, we have  $\ProjPj\ProjQk\ProjPjp\ProjQk\ProjPj=\KDelta{j'}{j}\ProjPj\ProjQk\ProjPj$. Then, for all $j'\neq\nobreak j$,
$\ProjPj\ProjQk\ProjPjp\ProjQk\ProjPj=\ProjPj\ProjQk\ProjPjp\ProjPjp\ProjQk\ProjPj%
 =\left(\ProjPjp\ProjQk\ProjPj\right)^{\dagger}\left(\ProjPjp\ProjQk\ProjPj\right)=0$,
so $\ProjPjp\ProjQk\ProjPj=0$ for all $j'\neq j$.%
\footnote{%
To obtain this result, \Luders introduced a lemma whose proof is obscurely incomplete: A projector is positive, but not positive-definite, so his ``scalar product'' is not a scalar product, so the Cauchy-Schwartz inequality may be applied only with further consideration of \emph{its} proof. Perhaps this is the reason that \citet[pp.~22-25]{Furry66}, in a lengthy (and rather inchoate) presentation of these theorems, merely refers to the \Luders paper rather than presenting this part of the proof. 
} %
Sum this over $j'$ to obtain $\ProjPj\ProjQk\ProjPj=\ProjQk\ProjPj$; take the hermitian conjugate and equate to obtain $\ProjPj\ProjQk=\ProjQk\ProjPj$.
\end{proof}

2. A measurement is said to be ``ignored'' if no action is taken based on its outcome; a probability calculation expresses an ignored measurement by the disjunction of all possible outcomes. \Luders showed that, if a preceding ignored measurement of $Q$ does not affect the probability of outcome of \p{j} in a succeeding measurement of $P$, then the projectors of all the \set{\q{k}} commute with the projector of \p{j}:
\begin{theorem}[Ignored measurement]\label{T:Luders2}
 \begin{equation}
  \sum_s\Pr[\sigma]{\q{s}\andthen\p{j}}=\Pr[\sigma]{\p{j}}\;\forall\sigma%
  \;\Longrightarrow\;\Proj{\p{j}}\Proj{\q{k}}=\Proj{\q{k}}\Proj{\p{j}}\;\forall k.
 \end{equation}
\end{theorem}
\begin{proof}
Translate the proposition to quantal terms to obtain $\sum_s\ProjQ_s\ProjPj\ProjQ_s=\ProjPj$.  Then, for each $k$, equate the two expressions obtained by multiplying on the left by $\ProjQk$ and by multiplying on the right by $\ProjQk$. 
\end{proof}

Because the disjunction of a complete set of propositions is identically ``true,'' it is generally believed that an ignored measurement may be ignored --- left out of the calculation entirely.  This, in fact, is the content of the formula of marginal probabilities: 
$\sum_s\Pr{\p{s}\wedge q}=\Pr{\Orj{s}{\p{s}}\wedge q}=\Pr{\text{true}\wedge q}=\Pr{q}$.
As noted by \citet{Margenau63a}, this is not generally the case in quantum mechanics. 
%: in general, the performance of an ignored measurement in a sequence of events cannot itself be ignored (unless it is the final event of the sequence). 
As \Luders shows here, it is exactly in the case of \emph{compatible} variables that the performance of an ignored measurement may be safely ignored.

\subsection{Davies' compatibility as time-order independence}

The order-independence of joint observation probability, $\Pr{\p{j}\andthen\q{k}}=\Pr{\q{k}\andthen\p{j}}$, leads to the commutability of the corresponding projectors. The was first proven by \citet[p.~15-17]{Davies76} under the requirements that the equality hold for all indices and that the projectors span the vector space. \citet{Kirkpatrick:Quantal} strengthened the theorem to refer to only a single value-pair, as presented here:

\begin{theorem}
\begin{equation}
  \Pr[\sigma]{\p{j}\andthen\q{k}}=\Pr[\sigma]{\q{k}\andthen\p{j}}\;\forall\sigma%
  \;\Longrightarrow\;%
  \Proj{\p{j}}\Proj{\q{k}}=\Proj{\q{k}}\Proj{\p{j}}.
\end{equation}
\end{theorem}  
\begin{proof}
Translate the proposition to quantal terms to obtain $\ProjPj\ProjQk\ProjPj=\ProjQk\ProjPj\ProjQk$. Define $\ProjC=\ProjPj\ProjQk-\ProjQk\ProjPj$; show $\ProjC^{\dagger}\ProjC=0$, hence $\ProjC=0$.
\end{proof}

\section{Definition of compatibility}\label{S:Definition}

We have three expressions available for a general definition of compatibility of values (\ie that lead, in quantum mechanics, to commuting projectors):
\begin{subequations}\begin{eqnarray}
 &\Prob[\sigma]{\p{j}}{\p{j}\andthen\q{k}}=1\quad\forall\sigma\label{E:NonDisturb}\\
 &\sum_s\Pr[\sigma]{\q{s}\andthen\p{j}}%
   =\Pr[\sigma]{\p{j}}\quad\forall\sigma\label{E:Ignored}\\
 &\Pr[\sigma]{\p{j}\andthen\q{k}}%
   =\Pr[\sigma]{\q{k}\andthen\p{j}}\quad\forall\sigma\label{E:OrderExchange}
\end{eqnarray}\end{subequations}

Within the quantal setting, of course, these three are equivalent (because each is equivalent with the commutability of the projectors). Which of these should be taken as the \emph{classical} definition of compatibility? 

Expression \eqref{E:OrderExchange} implies expression \eqref{E:NonDisturb} (replace $\sigma$ with $\sigma\andthen\p{j}$; use the repeatability of a \p{j}-filter: $\sigma\andthen\p{j}\andthen\p{j}=\sigma\andthen\p{j}$). If we assume expression \eqref{E:OrderExchange} for all $k$, it implies expression \eqref{E:Ignored} (sum over $k$, use the completeness of the \set{\q{k}}). It appears  that neither expression \eqref{E:NonDisturb} nor \eqref{E:Ignored} implies the other, nor does either imply expression \eqref{E:OrderExchange} (although I don't have counterexamples).

In quantum mechanics compatibility is a symmetric relation; this property is essential to the possibility of simultaneous measurability. This suggests that compatibility should be expressed in an explicitly symmetric way. 

Often compatibility is expressed in terms of the variables themselves, but two generally incompatible variables may have some simultaneously observable values. For this reason, it is preferable to express the criterion of compatibility in terms of a single pair of values. 

Expression \eqref{E:OrderExchange} involves only a single pair, and is the only one of the three which is symmetric. It contains the other two (``non-disturbance'') versions.  For these reasons, we consider it the best version, hence: 
\begin{definition*}[Compatible values]
The values \p{} and \q{} are compatible \IFF for every preparation state $\sigma$,
$\Pr[\sigma]{\p{}\andthen\q{}}=\Pr[\sigma]{\q{}\andthen\p{}}$. 
\end{definition*}

\section{Classical incompatibility}\label{S:Example}

Martin \citet[p.~275]{Strauss73a} pointed out that
``If we look back at the historical development of probability theory it must appear as a great misfortune that no stochastic game has ever been invented with more than one complete set of states \dots; if it had, the general theory of transition probabilities \dots\ would have been established \emph{before} the advent of quantum mechanics and the greater part of the discussions on the foundations of that theory would have been superfluous.'' Unfortunately, Strauss did not follow this insight with the development of such games; to the best of my knowledge, it was another thirty years before classical systems with incompatible variables appeared \citep{Kirkpatrick:Quantal, Kirkpatrick:ThreeBox}. 

In fact, however, such a stochastic game system \emph{is} in common use. A deck of playing cards forms the basis for games with two complete sets of states --- the \emph{Face} variable (with values Ace, 2, 3, \dots, King) and the \emph{Suit} variable (with values Clubs, Diamonds, Hearts, Spades). Generally, however, in textbook examples the drawing of cards is always done either \emph{with} replacement or \emph{without} replacement (``discard''); in either case the probability is independent of the order of occurrence,%
\footnote{%
This independence is obvious in the case of replacement, but a rather interesting arithmetic phenomenon in the case of discard.
} %
and incompatibility does not arise. However, there are many other replacement schemes which do not lead to this symmetry. The reader might find it interesting to construct such an example with, say, a deck of four cards, the King and the Queen of Hearts and of Spades, using the rule ``replace if \emph{Suit} is Spades, discard otherwise.'' The variables \emph{Face} and \emph{Suit} of this ``stochastic game'' fail the compatibility definition, expression \eqref{E:OrderExchange}. However, this is not a particularly compelling example to a physicist: the variables  are quite ``unphysical,'' not being repeatable (\eg having drawn a King, it is not certain that a King would be drawn next).  A little more is required to get variables which behave in a physically appropriate manner; we will find this in the following example (which, for variety, we express in a different physical form).

\subsection{A classical example exhibiting incompatibility and other ``quantal'' behaviors}

The system consists of a box of balls and an urn. The balls are of three different colors; some of the balls have white dots or stripes over the color: \ie each ball carries a value of the variables \Color (Yellow, Green, Blue) and  \Pattern (Plain, Dotted, Striped). The balls are otherwise all alike, particularly in mechanical properties (mass, diameter, elasticity, surface texture). Some of the balls are in the urn, the remainder in the box.

The value of the system's \Color variable is manifested by the following procedure: A ball is shaken out of the urn, and its \Color is noted and reported; all the balls (including those in the urn) are returned to the box, then all the balls of that reported \Color are put into the urn. The \Pattern is manifested in the analogous manner. 

We can prepare the system to have a particular \Color value by alternating manifestation of the \Pattern and of the \Color, repeating until the desired value of \Color is reported; systems so prepared are in the corresponding pure \Color p-state. (We prepare a pure \Pattern p-state analogously; initially, the urn is not empty.)

This classical probabilistic system provides \vN repeatability of observation in an otherwise random setting: $\Prob[\sigma]{q_j}{q_k}=\KDelta{j}{k}$ for $Q$ either \Color or \Pattern, and for all values (all $j$, $k$) of that variable. It also illustrates the reasonable and ordinary nature of the complementarity of the variables: after the manifestation of \Color, say, \Pattern has no value---there is no non-arbitrary way of assigning a value to \Pattern, no relevance of any such assignment to the ``physics'' of the system. 

Furthermore, as shown in the analysis (given in Appendix~B of \citet{Kirkpatrick:Quantal}) of the statistics of such systems, this system exhibits phenomena often thought of as quantal (summarized in Eqn.~(11) of that paper): in addition to variables incompatible in the sense defined in \RefSec{S:Definition}, it has no sharp-in-all-variables p-states, ignored observations may have statistical effects (the apparent failure of the marginal-probability formula), and interference may appear between alternative values (the apparent failure of logical distribution). Let us expand on this last point.

\subsection{Interference and incompatibility in this classical example} 
Consider a ``colorblind'' manifestation of \Color, wherein only Yellow and Grue (not-Yellow) are distinguished. The manifestation rule is followed literally --- if the color is not Yellow, then all Blue and all Green balls are placed in the urn. In every p-state the probability of Grue is the same as the sum of the probabilities of Green and Blue: $\Pr[\sigma]{\text{Grue}}=\Pr[\sigma]{\text{Blue}}+\Pr[\sigma]{\text{Green}}$. However, a preparation in a \Pattern p-state followed by a colorblind manifestation of \Color followed by a \Pattern manifestation shows interference between Grue and Blue-or-Green: for example,  $\Pr[\text{Plain}]{\text{Grue}\andthen\text{Dotted}}\neq \Pr[\text{Plain}]{\text{Blue}\andthen\text{Dotted}}+ \Pr[\text{Plain}]{\text{Green}\andthen\text{Dotted}}$. 

This is reflected in quantum mechanics by the fact that the superposition of several elementary p-states does not in itself imply anything ``unusual'' (\emph{pace} those who exercise themselves regarding macroscopic superposition). The unusual, that is, interference, appears only if the system prepared in a superposed p-state then undergoes a measurement of a variable incompatible with the values superposed; interference appears in the difference with the results expected from the disjunction of those values. (When the ``cross-term'' of a quantum expression is pointed out as being the ``interference'' terms, it is in fact exactly this difference between the expressions with and without complete manifestation. Such difference is basis-invariant; the cross-term, or rather the ability to point it out, is not.)

If the colorblind manifestation were followed by another \Color manifestation (or if the system had been prepared in a \Color p-state), there would be no interference between Grue and Blue-or-Green; incompatibility is essential to interference, in this classical example as well as in quantum mechanics.

\section{Conclusion}

Compatibility is a concept of classical probability, quite independent of quantum mechanics, and prior to quantum mechanics in every way but historically. This claim is justified by the existence of an expression in classical probability which, when combined with the quantal Hilbert-space formalism, is exactly equivalent with the commuting-operator definition, and which may be exemplified in non-quantal probability systems.

Incompatible variables in classical-probability systems exhibit a wide range of properties generally thought of as quantal: the impossibility of filtering to a fully sharp state (hence the failure of an ensemble picture), complementarity (``non-reality''), interference (``nondistributive logic'')--- much interpretive concern has been expended over the years on such chimera. (Some) physicists have been glorying in the incomprehensibility of the world described by quantum mechanics since the beginning; we need now to bring a broader vision to bear --- there is more to our toolkit than the models and ideas of deterministic classical physics.

%%%%%%%%%%%%%%%%%%%%%%%%%%%%%%%%%%%%%%%%%%%%%%%%%%%%%%%%%%%%%%%%%%%%%%%%%%%%%%%%%%%%%%%%
\clearpage%%LO

\bigskip\bigskip\noindent\textbf{Acknowledgments}\bigskip\\*
I which to thank Claudio Garola and Sandro Sozzo (personal communication; see also Section 3.2 of their (\citeyear{GarolaSozzo03})) for bringing to my attention Davies' probabilistic definition of compatibility. 

%%%%%%%%%%%%%%%%%%%%%%%%%%%%%%%%%%%%%%%%%%%%%%%%%%%%%%%%%%%%%%%%%%%%%%%%%%%%%%%%%%%%%%%%
%% BIBLIOGRAPHY %%%%%%%%%%%%%%%%%%%%%%%%%%%%%%%%%%%%%%%%%%%%%%%%%%%%%%%%%%%%%%%%%%%%%%%%
%%%%%%%%%%%%%%%%%%%%%%%%%%%%%%%%%%%%%%%%%%%%%%%%%%%%%%%%%%%%%%%%%%%%%%%%%%%%%%%%%%%%%%%%
 \renewcommand{\refname}{\sc References}%xxx
 \footnotesize%
%%%%%%%%%%%%%%%%%%%%%%%%%%%%%%%%%%%%%%%%%%%%%%%%%%%%%%%%%%%%%%%%%%%%%%%%%%%%%%%%%%%%%%%
% \bibliographystyle{myapalike}
% \bibliography{JAbbrevs,QM}%\NotesBib}

%%%%%%%%%%%%%%%%%%%%%%%%%%%%%%%%%%%%%%%%%%%%%%%%%%%%%%%%%%%%%%%%%%%%%%%%%%%%%%%%%%%%%%%%
%%% END DOCUMENT %%%%%%%%%%%%%%%%%%%%%%%%%%%%%%%%%%%%%%%%%%%%%%%%%%%%%%%%%%%%%%%%%%%%%%%
%%%%%%%%%%%%%%%%%%%%%%%%%%%%%%%%%%%%%%%%%%%%%%%%%%%%%%%%%%%%%%%%%%%%%%%%%%%%%%%%%%%%%%%%
\end{document}